# Fracture Toughness of Silicate Glasses: Insights from Molecular Dynamics Simulations

Yingtian Yu,[1] Bu Wang,[1] Young Jea Lee,[1] and Mathieu Bauchy[1]

[1]Department of Civil and Environmental Engineering, University of California, Los Angeles, CA 90095, United States

## ABSTRACT

Understanding, predicting and eventually improving the resistance to fracture of silicate materials is of primary importance to design new glasses that would be tougher, while retaining their transparency. However, the atomic mechanism of the fracture in amorphous silicate materials is still a topic of debate. In particular, there is some controversy about the existence of ductility at the nano-scale during the crack propagation. Here, we present simulations of the fracture of three archetypical silicate glasses using molecular dynamics. We show that the methodology that is used provide realistic values of fracture energy and toughness. In addition, the simulations clearly suggest that silicate glasses can show different degrees of ductility, depending on their composition.

## INTRODUCTION

Brittleness is the main limitation of glasses, as impacts, scratches or vibrations can result in undesirable or even dangerous fracture. Indeed, glasses lack a stable shearing mechanism, thus showing very poor ductility and, consequently, high brittleness [1, 2]. This is a serious safety concern, as the number of injuries related to glass (e.g., during car crashes or by broken bottles) is significant. Further, improving the mechanical properties of glasses is crucial to address major challenges in energy, communication and infrastructure arenas [3]. For example, strength, toughness and stiffness are a major bottleneck for further development of short-haul high-capacity telecommunication and fiber-to-the-home technologies, flexible substrates and roll-to-roll processing of displays, solar modules, planar lighting devices, the next generation of touch screen devices, large scale and high altitude architectural glazing, ultra-stiff composites and numerous other applications. Increasing the strength and toughness of glass would not only enable new applications, but also lead to a significant reduction of material investment for existing applications while achieving comparable performances [4].

To improve the ductility of glasses, current techniques focus on compositing [5], inclusion of holes [6] or surface treatments [7]. However, these treatments often result in undesirable side effects such as a loss of transparency [3]. An alternative option is to enhance the intrinsic ductility of glasses by tuning their atomic topology, which is mainly a function of their composition. Such intrinsic optimization, which has been established as a Grand Challenge for glass [4], is the focus of the present study. Fulfillment of this goal requires elucidation of the atomistic mechanism of fracture in glasses. Indeed, although glasses are typically brittle materials at the macro-scale, there remains some controversy about the existence of ductility at the nano-scale. Hence, as opposed to an ideal brittle fracture model, in which cracks would propagate based on a series of chemical bond rupture events [8], it has been suggested that oxide glasses should show plastic deformations at the vicinity of the crack tip [9], although this is still a matter of debate [10, 11].

Here, relying on molecular dynamics simulations and well-established inter-atomic potentials, we present a general methodology [12] allowing us to compute the fracture toughness and critical energy release rate of glassy silica (S), sodium silicate (NS), and calcium aluminosilicate (CAS). On the other hand, the computation of their surface energy enables to quantify their relative brittleness.

**SIMULATION DETAILS**

**Preparation of the glasses**

To assess the ability of the molecular dynamics simulation to predict realistic values of fracture toughness and critical energy release rate, we focus on three different silicate glasses: (1) pure silica, the structural basis for all silicate glasses, (2) a sodium silicate glass of composition $(Na_2O)_{0.30}(SiO_2)_{0.70}$, a base material for alkali silicate industrial glass strengthened by ion-exchange, and (3) a calcium aluminosilicate glass of composition $(SiO_2)_{0.60}(Al_2O_3)_{0.10}(CaO)_{0.30}$, the basis for all alkali-free display glasses. Based on these glasses, we aim to understand the effects of the depolymerization of the silica network by alkali atoms and of the inclusion of intermediate network formers species like calcium atoms.

For this study, we relied on the following well-established inter-atomic potentials. For silica, we used a modified BKS potential [17, 18], which has been found to offer a realistic mechanism for the mode I failure of silica [19]. The sodium silica glass was simulated using a Buckingham potential parameterized by Teter [20], which has been shown to provide excellent results for structure, dynamics, and mechanics [21–25]. Finally, the calcium aluminosilicate glass was prepared using the potential of Matsui [26], reparametrized by Jakse et al [27], based on ab initio calculations [28]. The ability of this potential to predict a realistic structure and good mechanical properties has recently been reported [29].

These three glasses have been prepared in a consistent way, with the LAMMPS package [30], using an integration time-step of 1 fs. Coulomb interactions were evaluated by the Ewald summation method, with a cutoff of 12 Å. The short-range interaction cutoff was chosen at 8.0 Å for NS and CAS, and at 5.5Å for S. A liquid slab made of around 18000 atoms were first generated by placing the atoms randomly in the simulation box. The system was then equilibrated at 5000 K in the NPT ensemble for 1 ns, at zero pressure, to assure the loss of the memory of the initial configuration. Glasses were formed by linear cooling of the liquids from 5000 to 300 K, with a cooling rate of 1 K/ps. Once formed, glasses were relaxed at zero pressure and 300 K for 1 ns in the NPT ensemble.

**Validation of the predicted structure**

As a preliminary step, we checked that the used potentials offer a realistic structure for the three selected glasses by computing the total pair distribution function (PDF) and the neutron structure factor, and compare them to available experimental data. The partial structure factors were first calculated from the pair distribution functions (PDF) $g_{ij}(r)$:

$$S_{ij}(Q) = 1 + \rho_0 \int_0^R 4\pi r^2 (g_{ij}(r) - 1) \frac{\sin(Qr)}{Qr} F_L(r) dr \qquad (1)$$

where $Q$ is the scattering vector, $\rho_0$ is the average atom number density and R is the maximum value of the integration in real space (here R = 16 Å). The $F_L(r) = sin(\pi r/R)/(\pi r/R)$ term is a Lorch-type window function, used to reduce the effect of the finite cutoff of r in the integration [31]. As discussed in Ref. [32], the use of this function reduces the ripples at low $Q$, but induces a broadening of the structure factor peaks. The total neutron structure factor can then be evaluated from the partial structure factors following:

$$S_N(Q) = (\sum_{i,j=1}^{n} c_i c_j b_i b_j)^{-1} \sum_{i,j=1}^{n} c_i c_j b_i b_j S_{ij}(Q) \quad (2)$$

where $c_i$ is the fraction of specie $i$ (Si, O, Na, Al, or Ca) and $b_i$ is the neutron scattering length of the species (given by 4.149, 5.803, 3.63, 3.449, and 4.700 fm for silicon, oxygen, sodium, aluminum, and calcium atoms, respectively [33]).

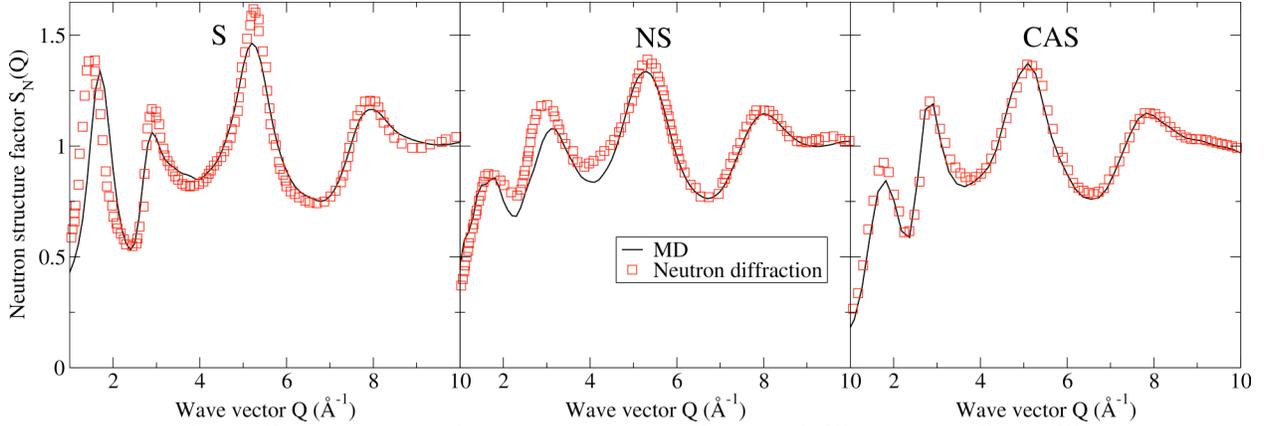

**Figure 1.** (Color online) Computed neutron structure factor of silica (S), sodium silicate (NS), and calcium aluminosilicate (CAS), compared with neutron diffraction measurements [13–16].

Figure 1 shows the computed neutron structure factors, each of them compared with data from neutron scattering [13–16]. We note that the experimental structure factors are fairly well reproduced both at low and high $Q$, which suggest a good agreement between simulated and experimental structures both at the medium- and short-range order. Overall, the present level of agreement is comparable to that obtained in previous studies using the same inter-atomic potentials [34–36].

We now compare the predicted structure with experimental data in real space. Indeed, as claimed by Wright [37], real space and reciprocal space correlation functions, respectively, emphasize different features of a given structure. Hence, it is necessary to compare the simulation to experiments in both spaces. Coming back to real space, the total PDFs $g(r)$ were calculated from the partials:

$$g(r) = (\sum_{i,j=1}^{n} c_i c_j b_i b_j)^{-1} \sum_{i,j=1}^{n} c_i c_j b_i b_j g_{ij}(r) \quad (3)$$

and compared to experimental data [13–16]. The latter were obtained via the Fourier transform of the experimental neutron structure factor, using the previously mentioned Lorch-type window function to reduce the ripples at low $r$. To take into account the maximal scattering vector $Q_{max}$ of the experimental structure factor, the computed $g(r)$ was broadened by following the methodology described by Wright [37].

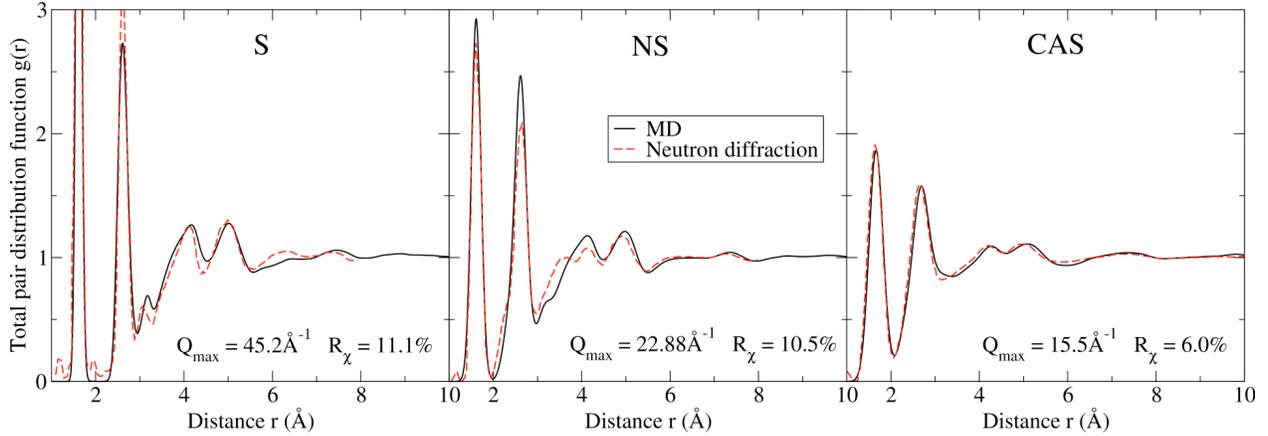

**Figure 2.** (Color online) Computed total pair distribution functions of silica (S), sodium silicate (NS), and calcium aluminosilicate (CAS), compared with neutron diffraction measurements [13–16].

Figure 2 shows the computed total PDFs, compared with neutron diffraction measurements [13–16]. Once again, we observe a fairly good agreement between the simulated and experimental PDFs, both at low and high $r$. Rather than relying on a simple visual observation, we quantified the agreement between experimental and simulated correlation functions by calculating Wright's $R_\chi$ factor:

$$R_\chi = \left[ \frac{\sum_{i=1}^{n}(g(r) - g_{ref}(r))^2}{\sum_{i=1}^{n}(g_{ref}(r))^2} \right] \qquad (4)$$

where $g_{ref}(r)$ is the experimental total PDF. These factors, calculated over the range in $r$ from 1.0 Å to 8.0 Å, are given in figure 2 and provide a quantitative measurement of the relative quality of the used potentials.

**Simulations of fracture**

Recently, Brochard et al. [12] introduced a new method to study fracture properties at the smallest scales, based on molecular dynamics simulations. This approach relies on the energetic theory of fracture mechanics [38–40] and consists of thermodynamic integration during crack propagation. This method does not involve any assumption about the mechanical behavior of the material during the fracture and can thus capture fracture properties of brittle as well as ductile systems [12].

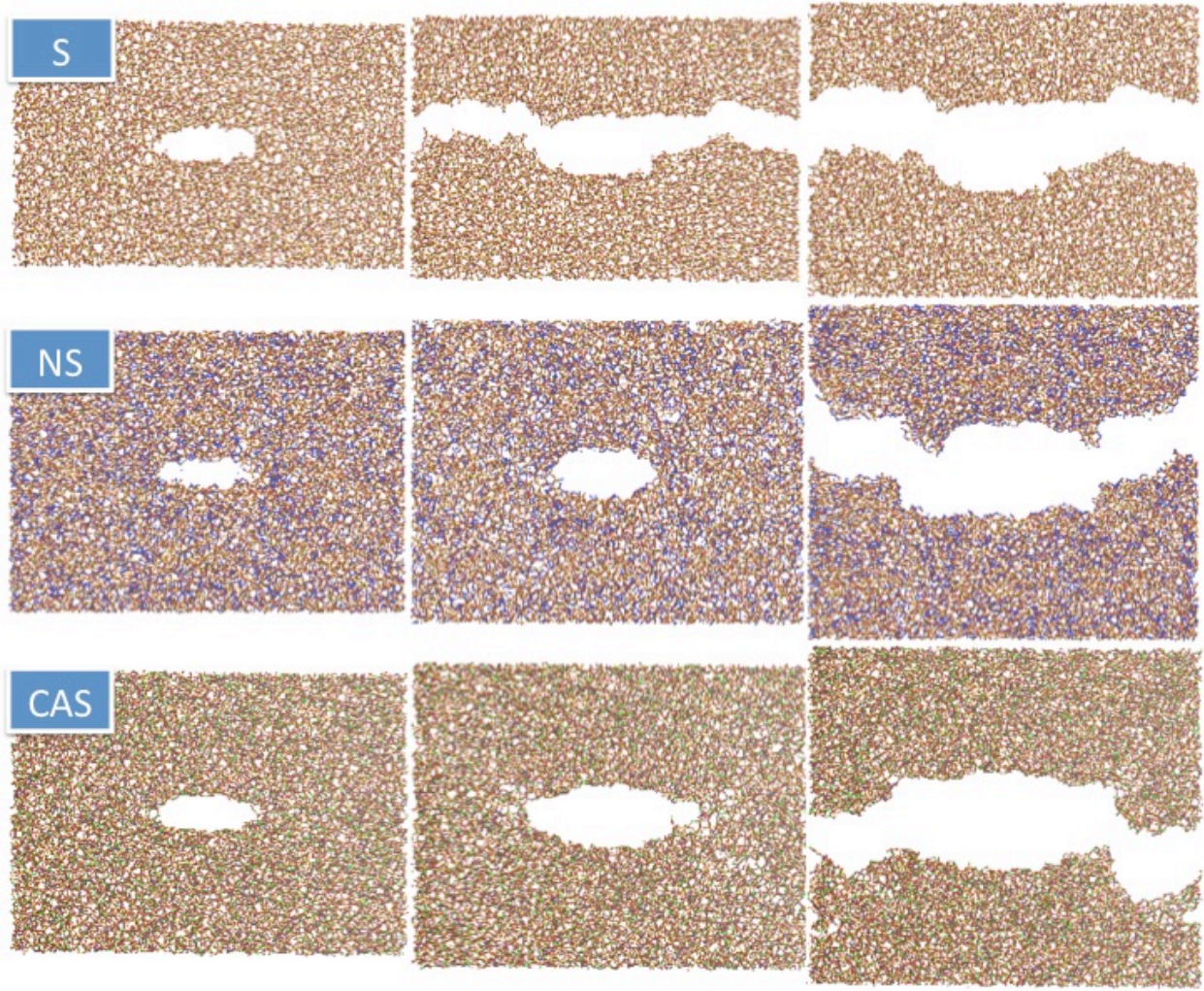

**Figure 3.** (Color online) Snapshots of the atomic configurations of silica (S), sodium silicate (NS), and calcium aluminosilicate (CAS), with strains of 0.09, 0.18, and 0.30, respectively. Silicon, aluminum, calcium, sodium, and oxygen atoms are represented in yellow, grey, green, blue, and red, respectively.

In the following, we focus on fracture in mode I, i.e., with an opening mode and a loading normal to the crack plane. As illustrated by figure 3, a crack is first initiated into the molecular sample. Such cracks are expected to exist naturally in real materials. Note that, as in experiments, starting from a pre-cracked system is necessary to perform fracture toughness measurements. The initial crack is created by removing atoms located inside an elliptic volume along the $x$ direction. The ellipse is chosen to be five times larger in the $y$ direction than in the $z$ direction, thus inducing a strong concentration of the stress at the crack tips. Its length is chosen to be around 50 Å, slightly adjusted in each case to assure a neutral system. Note that the initial length must be long enough for the initial hole to be stable but small as compared with the box length in the $y$ direction (the typical system size is $16 \times 150 \times 100$ Å in the $x, y, z$ directions, respectively).

Before any tension is applied, the system is fully relaxed to be unstressed; thus, its mechanical energy $P$, involved by strain, becomes zero. The procedure then consists of increasing the size $L_z$ of the system in the direction orthogonal to the initial crack until its full propagation along the $y$ axis. $L_z$ is incremented stepwise by 1% of its initial unstressed value $L_{z0}$ up to $L_{zmax} = 1.5 L_{z0}$. After each increase of the tensile strain $\varepsilon = (L_z - L_{z0})/L_{z0}$, the system is relaxed for 50 ps before performing a statistical averaging stage for another 50 ps. During the latter phase, the stress in the $z$ direction $\sigma_z$ is computed with the virial equation [41].

Note that the entire fracture simulation is operated within the canonical NVT ensemble, in which the temperature is controlled by a Nose–Hoover thermostat [42, 43]. Hence, we are unable to capture potential heat transfers during the fracture. In fact, this procedure has not been designed to model the kinetics of crack propagation. On the contrary, thermodynamic quantities are always integrated when the system is at equilibrium, at each strain step. The phonons that arise during the fracture are annealed by the thermostat and, therefore, are not included in the following thermodynamic integration.

As the crack starts to propagate, some elastic energy $P$ is released to create new surface. This is captured by the energy release rate $G$:

$$G = -\frac{\partial P}{\partial A} \tag{5}$$

where $A$ is the crack area. When propagation occurs, the energy release rate is equal to the critical energy release rate $G_c$, which is considered as a property of the material. Once the crack propagation is complete, the system becomes unstressed again, so that $P = 0$, the mechanical energy having been released by crack propagation. The integration of $\sigma_z$ over the whole process, i.e., the external work, thus provides the critical energy release rate $G_c$:

$$G_c = \frac{\Delta F}{\Delta A} = \frac{L_x L_y}{\Delta A_\infty} \int_{L_{z0}}^{L_{zmax}} \sigma_z dL_z \tag{6}$$

where $F$ is the free energy of the system and $\Delta A_\infty = A_\infty - A_0$ is the total area of surface created at the end of the fracture, when the crack has fully propagated. This formula is a direct consequence of Griffith theory of fracture [38]. It is worth noting that evaluating the crack area at the end of the fracture may not be straightforward as the created surface may show some roughness. To make an accurate estimate of the critical energy release rate, the real surface area has been calculated using the procedure proposed in Ref. [12].

Alternatively to the energetic approach, the notion of fracture toughness $K_{Ic}$ is usually used in engineering application. This quantity was introduced by Irwin [44] as the maximum stress intensity at the crack tip a solid can undergo, and below which propagation cannot occur. The relationship between $K_{Ic}$ and $G_c$ is given by the Irwin formula [44]:

$$G_c = H_I K_{Ic}^2 \tag{7}$$

where $H_I$ is given in Ref. [45] for transversely isotropic solids and can be written in terms of the stiffness constants $C_{ij}$, using Voigt notation, as:

$$H_I = \frac{1}{2}\sqrt{\frac{C_{11}}{C_{11}C_{33} - C_{13}^2}\left(\frac{1}{C_{44}} + \frac{2}{C_{13} + \sqrt{C_{11}C_{33}}}\right)} \qquad (8)$$

in plane strain, as is the case of the current study. Note that, although we rely on a general energetic approach that does not assume a purely brittle fracture, we keep in mind that the relation between $G_c$ and $K_{Ic}$ was derived in the context of Linear Elastic Fracture Mechanics (LEFM). The full elastic tensor $C_{ij}$ was computed for a bulk system, before the introduction of the initial crack. The elements of the stiffness tensor are obtained by calculating the curvature of the potential energy $U$ with respect to small strain deformations $\acute{U}_i$ [25]:

$$C_{ij} = \frac{1}{V}\frac{\partial^2 U}{\partial \acute{U}_i \partial \acute{U}_j} \qquad (9)$$

where $V$ is the volume of the system. In isotropic materials, which is the case of the present glasses, Eq.7 reduces to the usual Irwin formula [39]:

$$G_c = \frac{1-v^2}{E}K_{Ic}^2 \qquad (10)$$

where $E$ is the Young's modulus. Hence, this method provides an indirect computation of $K_{Ic}$, by using a purely energetic approach. The results obtained for the three considered glasses are then presented and compared to available experimental data.

**RESULTS**

Figure 4 shows the computed stress-strain curves (stress $\sigma_z$ with respect to the tensile strain $\varepsilon$) for the three glasses. At low strain (up to 12%, 9%, and 18 % for S, NS, and CAS, respectively), the mechanical response is fairly linear elastic. The stress thus increases linearly with the strain, up to around 9 GPa, with the slope related to the Young's modulus of the system. During this stage, which is observed for the three glasses, the crack does not propagate and the free energy of the system is stored in the form of mechanical elastic energy only.

At larger strain, the crack starts to propagate. As shown in figure 4, silica is characterized by a brittle fracture, as the crack suddenly propagates above a critical strain of 13 %. This manifests by a drop of the tensile stress to zero, which is comparable to what has been observed for quartz [12]. On the contrary, NS and CAS glasses break in more ductile way, in the sense that the crack does not propagate instantly after a given critical strain. Thanks to their internal flexibility, their networks rather deform to prevent the fracture from occurring, as observed in the snapshots inside figure 3. The glasses eventually break at 26 % and 28 % for NS and CAS, respectively.

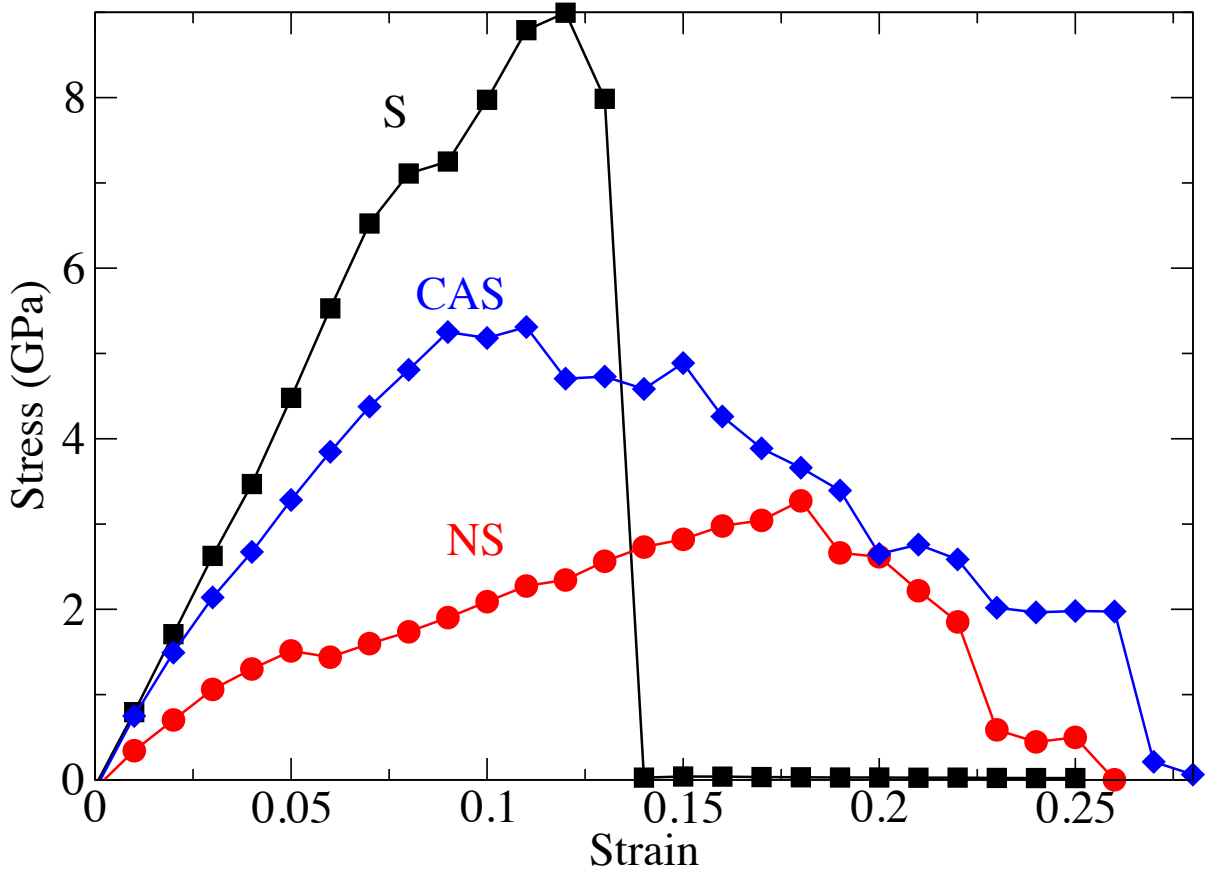

**Figure 4.** (Color online) Computed stress as a function of the tensile strain imposed to the system, for silica (S), sodium silicate (NS), and calcium aluminosilicate (CAS) glasses, respectively.

The ductile behaviors that are observed for NS and CAS require an extra care: indeed, as the crack propagates, irreversible processes, such as plasticity, occur inside a process zone around the crack tip. An estimated length of this plasticity zone $r_{pl}$ can be evaluated using the Dugdale–Barenblatt formula [46–48]:

$$r_{pl} = \frac{\pi}{8}\left(\frac{K_{Ic}}{\sigma_{pl}}\right)^2 \qquad (11)$$

where $\sigma_{pl}$ is the plastic yield stress of the material.
At the end of the fracture, the process zones located at both sides of the crack eventually overlap because of the periodic boundary conditions. As suggested in Ref. [12], this feature can be taken into account by replacing in Eq. 6 the real crack area $\Delta A_\infty$ by an effective area given by $\Delta A_{\infty,eff} = \Delta A_\infty - L_x r_{pl}/2$.

The final values of the fracture energy and toughness, after all the correction have been made, are reported in table I. Although it is known that measured values of fracture toughness are very sensitive to the method used, the preparation of the glass, and the environment (dry or in presence of water), we obtain a surprisingly good agreement between computed and experimental data for the three glasses. Note that the CAS glass show a lower fracture energy than that of NS, but a higher fracture toughness, which results from its higher Young's modulus.

## DISCUSSION

We now aim to quantify the atomic-scale relative brittleness of the three glasses. To this end, the critical energy release rate $G_c$ can be expressed as follows:

$$G_c = G_{el} + G_{diss} \tag{12}$$

where $G_{el}$ is the elastic contribution to the fracture energy, i.e., arising from the stress accumulated in the linear-elastic regime, and $G_{diss}$ captures all forms of dissipated energy linked to irreversible processes and would be equal to zero for a perfectly brittle material. $G_{el}$ was evaluated by integrating the stress-strain curves up to the strain at which the maximum stress is obtained. This allows us to quantify the ductility of each by computing a brittleness parameter $B = G_{el}/G_c$, which is equal to 1 for a perfectly brittle material. The computed values of $B$ are reported in Tab. I. We observe that, with such a definition of the brittleness, none of the considered glass is perfectly brittle at the atomic scale. If pure silica is the closest to show an ideal brittleness, NS and CAS clearly show a high ductility.

Although glasses are typically brittle materials at the macro-scale, there remains some controversy about the existence of ductility at the nanoscale. Hence, as opposed to an ideal brittle fracture model, in which cracks would propagate based on a series of chemical bond rupture events [8], it has been suggested that oxide glasses should show plastic deformations at the vicinity of the crack tip [9], although this is still a matter of debate [10, 11]. The intrinsic brittleness or ductility of glass appears to strongly depend on the composition and the structure, and has recently been shown to be correlated to the Poisson's ratio [19].

**Table I.** Computed fracture energy ($G_c$), elastic fracture energy ($G_{el}$), fracture toughness ($K_{Ic}$), and brittleness index ($B$) for silica (S), sodium silicate (NS) and calcium aluminosilicate (CAS). Experimental values are added in parenthesis, when available.

| Glass | $G_c$ (J/m$^2$) | $K_{Ic}$ (MPa.m$^{1/2}$) | $B$ |
|---|---|---|---|
| S | 9.2 (9.0±0.4 [49]) | 0.81 (0.81±0.02 [49]) | 0.88 |
| NS | 6.4 (7±1 J/m$^2$ [50]) | 0.64 | 0.73 |
| CAS | 4.3 | 0.66 (0.63 ±0.5 J/m$^2$ [51]) | 0.38 |

## CONCLUSIONS

A selection of three silicate glasses has been simulated by molecular dynamics in order to evaluate their fracture energy, fracture toughness and relative brittleness. We observe a good agreement between computed values and available experimental data, although it is unclear how much the glass preparation, experimental setup and environment have affected the measured values of fracture toughness. Despite these uncertainties, this methodology was shown to provide realistic trends of fracture toughness and energies, especially with respect to composition, for a given family of glass. This allows the details of such relationships to be further investigated in the future.